\magnification1200
\parindent0pt
\parskip3pt
\baselineskip=1.5\baselineskip

\def\estwz{\tilde\theta_{\alpha\beta}}
\def\estlq{\hat\theta_{\alpha\beta}}

\centerline{\bf A comment on}
\centerline{\bf ``New non-parametric inferences for low-income proportions''}
\centerline{\bf by Shan Luo and Gengsheng Qin}

\vskip1truecm
\centerline{Wojciech Zieliñski}
\centerline{Department of Econometrics and Statistics}
\centerline{Warsaw University of Life Sciences}
\centerline{Nowoursynowska 159, 02-776 Warszawa}
\centerline{e-mail: wojciech\_zielinski@sggw.pl}
\centerline{http://wojtek.zielinski.statystyka.info}

\bigskip

{\bf Abstract.} Shan Luo and Gengsheng Qin published the article "New non-parametric inferences for low-income proportions" Ann Inst Stat Math, 69, 599-626. In the note their approach is compared to Zieliñski 2009 approach.

\bigskip


\bigskip

Low-income proportion (LIP) known also as At-Risk-of-Poverty-Rate (ARPR) is an important index in describing the inequality of an income distribution. It is often used to evaluate the social economic and poverty status of a population. A low-income proportion is defined as the proportion of the population income below a given fraction $\alpha$ ($0 < \alpha < 1$) of the $\beta$ ($0 < \beta < 1$) quantile of an income distribution. Let $X\in[0,\infty)$ be an income variable with cumulative distribution function $F(\cdot)$ and denote $\xi_\beta$ as the $\beta$ quantile $F^{-1}(\beta)$ of $F$. Then, the $\alpha$ fraction of the $\beta$ quantile ƒ$\alpha\xi_\beta$ is the income line, and the low-income proportion (At-Risk-of-Poverty-Rate) is
$$\theta_{\alpha\beta}=F(\alpha\xi_\beta).$$
Let $X_1,\ldots,X_n$ be a sample of disposable incomes of randomly drawn $n$ persons and let $\hat\xi_{\beta}$ denote the estimator of $\xi_{\beta}$. The natural estimator $\estwz$ is defined as
$$\estwz={1\over n}\#\{X_i\leq\alpha\cdot\tilde\xi_{\beta}\},$$
where $\#S$ denotes the cardinality of the set $S$. This estimator may be equivalently written as
$$\estwz={1\over n}\max\left\{j:X_{j:n}\leq\alpha X_{\lfloor\beta n\rfloor+1:n}\right\},\eqno{(W)}$$
where $X_{1:n}\leq\cdots\leq X_{n:n}$ is the ordered sample and $X_{\lfloor\beta n\rfloor+1:n}$ is the estimator of $\xi_{\beta}$, i.e. $\tilde\xi_{\beta}=X_{\lfloor\beta n\rfloor+1:n}$.

Luo and Qin proposed the following estimator of $\theta_{\alpha\beta}$:
$$\estlq=F_n(\alpha\hat\xi_{\beta}),\eqno{(LQ)}$$
where $F_n(\cdot)$ is the empirical distribution function and $\hat\xi_{\beta}=F_n^{-1}(\beta)$. Note that if $F_n$ is the classical empirical distribution function than $(LQ)$ and $(W)$ are the same.

Luo and Qin made an asymptotic consideration. In their approach they firstly estimate the cumulative distribution function $F$ using a kernel estimator and then investigate the properties of the obtained estimator. By making use of the asymptotic normality of $\estlq$ the confidence interval for $\theta_{\alpha\beta}$ was constructed.

Zieliñski R. (2006) showed that $\estwz$ is almost unbiased even for small sample sizes. He also calculated its variance. Zieliñski W. (2009) showed that the random variable $\eta=\#\{X_i\leq\alpha\cdot\tilde\xi_{\beta}\}$ is distributed as (almost) binomial with parameters $M=\lfloor\beta n\rfloor$ and $\theta_{\alpha\beta}/\beta$ (if $F$ is a power distribution function it is exactly binomial). The confidence interval for $\theta_{\alpha\beta}$ (at the confidence level $\gamma$) has the form (Clopper and Pearson 1934)
$$\left(qB^{-1}\left(\eta,M-\eta+1;{1-\gamma\over2}\right);qB^{-1}\left(\eta+1,M-\eta;{1-\gamma\over2}\right)\right),\eqno{(*)}$$
where $B^{-1}(a,b;\delta)$ is the $\delta$ quantile of beta distribution with parameters $(a,b)$.

The coverage probability of $(*)$ (for $\alpha=0.5$, different $\beta$ and $\gamma=0.95$) calculated (not simulated) for chi-square $\chi^2_3$ and lognormal $\log N(0,1)$ distributions are given in Tables 1 and 2, respectively. Results are similar to those obtained in simulations by Luo and Qin.

$$\vbox{\tabskip5em minus4.9em\offinterlineskip \halign to\hsize{
\strut\hfil$#$&#\vrule&&\hfil$#$\hfil&\hfil$#$\hfil&#\vrule\cr
\multispan{13}{\bf Table 1.} Coverage probability and the length of $(*)$ for $F=\chi^2_3$\hfil\cr\noalign{\vskip5pt}
 &&\multispan2\hfil$\beta=0.5$\hfil&&\multispan2\hfil$\beta=0.6$\hfil&&\multispan2\hfil$\beta=0.7$\hfil&&\multispan2\hfil$\beta=0.8$\hfil\cr
 &&\multispan2\hfil$\theta_{\alpha\beta}=0.2429$\hfil&&\multispan2\hfil$\theta_{\alpha\beta}=0.3115$\hfil&&\multispan2\hfil$\theta_{\alpha\beta}=0.3921$\hfil&&\multispan2\hfil$\theta_{\alpha\beta}=0.4915$\hfil\cr
n&&covarage&length&&covarage&length&&covarage&length&&covarage&length\cr\noalign{\hrule}
 500&&0.9523&0.0605&&0.9482&0.0658&&0.9394&0.0699&&0.9290&0.0725\cr
 800&&0.9494&0.0475&&0.9468&0.0517&&0.9400&0.0551&&0.9280&0.0570\cr
1000&&0.9509&0.0424&&0.9412&0.0459&&0.9427&0.0493&&0.9293&0.0510\cr\noalign{\vskip10pt}
}}$$
$$\vbox{\tabskip5em minus4.9em\offinterlineskip \halign to\hsize{
\strut\hfil$#$&#\vrule&&\hfil$#$\hfil&\hfil$#$\hfil&#\vrule\cr
\multispan{13}{\bf Table 2.} Coverage probability and the length of $(*)$ for $F=\log N(0,1)$\hfil\cr\noalign{\vskip5pt}
 &&\multispan2\hfil$\beta=0.5$\hfil&&\multispan2\hfil$\beta=0.6$\hfil&&\multispan2\hfil$\beta=0.7$\hfil&&\multispan2\hfil$\beta=0.8$\hfil\cr
 &&\multispan2\hfil$\theta_{\alpha\beta}=0.2441$\hfil&&\multispan2\hfil$\theta_{\alpha\beta}=0.3300$\hfil&&\multispan2\hfil$\theta_{\alpha\beta}=0.4330$\hfil&&\multispan2\hfil$\theta_{\alpha\beta}=0.5590$\hfil\cr
n&&covarage&length&&covarage&length&&covarage&length&&covarage&length\cr\noalign{\hrule}
 500&&0.9368&0.0595&&0.9316&0.0645&&0.9134&0.0666&&0.8984&0.0664\cr
 800&&0.9345&0.0468&&0.9233&0.0503&&0.9165&0.0526&&0.9039&0.0524\cr
1000&&0.9366&0.0418&&0.9288&0.0452&&0.9140&0.0469&&0.8984&0.0466\cr
}}$$

For ARPR defined in Eurostat document (Doc. IPSE/65/04/EN page 11) i.e. for $\alpha=0.6$ and $\beta=0.5$ Zieliñski W. (2009) calculated coverage probabilities for different parent distributions and sample sizes $20$, $200$ and $2000$. Those results show that the coverage probability of $(*)$ is near the nominal confidence level.

The estimator $\estwz$ may be recommended in applications because of its simplicity and good properties in small as well as in large samples. The estimator $\estlq$ is more complicated and is applicable only for large sample sizes.

\bigskip
{\bf References}
\font\kapitaliki=plcsc10
\def\art#1#2#3#4#5{\noindent\hangindent=0.5truecm \hangafter=1 {\kapitaliki #1}\ (#2):\ ``#3,''\ {\it #4}, #5.\par}

\art{Luo, S., Qin, G.}{2017}{New non-parametric inferences for low-income proportions}{Ann Inst Stat Math}{69, 599-626, DOI 10.1007/s10463-016-0554-0}

\art{Clopper, C. J., Pearson, E. S.}{1934}{The Use of Confidence or Fiducial Limits Illustrated in the Case of the Binomial}{Biometrika}{26, 404-413}

\art{Zieliñski, R.}{2006}{Exact distribution of the natural ARPR estimator in small samples from infinite populations}{Statistics in Transition  new series}{7, 881-888}

\art{Zieliñski, W.}{2009}{A nonparametric confidence interval for At-Risk-of-Poverty-Rate}{Statistics in Transition new series}{10, 437-444}

\bye